\documentclass[12pt]{article}

\usepackage{scicite}
\usepackage{times}
\usepackage{changes}
\usepackage{color}
\usepackage{amsfonts}
\usepackage[english]{babel}
\usepackage{subfig} 
\usepackage{graphicx}
\usepackage{epsf} 
\usepackage{amsmath,amssymb}

\definecolor{MATblue}{rgb}{0 .4470 .7410}
\definecolor{MATorange}{rgb}{.850 .325 .099}
\definecolor{MATyellow}{rgb}{.929 .694 .125}
\definecolor{MATpurple}{rgb}{.494 .184 .556}

\newcommand{\ket}[1]{|#1\rangle}
\newcommand{\bra}[1]{\langle #1|}
\newcommand{\braket}[2]{\langle #1|#2\rangle}

\newcommand{\ketbradif}[2]{\ket{#1}\bra{#2}}
\newcommand{\ketbra}[1]{\ketbradif {#1}{#1}}

\newcommand\numberthis{\addtocounter{equation}{1}\tag{\theequation}}

\newenvironment{sciabstract}{%
	\begin{quote} \bf}
	{\end{quote}}

\title{Precision measurement of atomic isotope shifts using a two-isotope entangled state}

\author
{Tom Manovitz,$^{\ast}$ Ravid Shaniv, Yotam Shapira,\\
	Roee Ozeri and Nitzan Akerman\\
	\\
	\normalsize{Department of Physics of Complex Systems, Weizmann Institute of Science,}\\
	\normalsize{234 Herzl St, Rehovot, 7610001 Israel}\\
	\\
	\normalsize{$^\ast$To whom correspondence should be addressed; E-mail:  tom.manovitz@weizmann.ac.il.}
}

\date{March 28th, 2019}

\begin{document} 
	
	
	\baselineskip24pt
	
	
	\maketitle

	\begin{sciabstract}
Atomic isotope shifts (ISs) are the isotope-dependent energy differences in the atomic electron energy levels. These shifts serve an important role in atomic and nuclear physics, and particularly in the latter as signatures of nuclear structure. Recently ISs have been suggested as unique probes of beyond Standard Model (SM) physics, under the condition that they be determined significantly more precisely than current state of the art. In this work we present a simple and robust method for measuring ISs with ions in a Paul trap, by taking advantage of Hilbert subspaces that are insensitive to common-mode noise yet sensitive to the IS. Using this method we evaluate the IS of the $5S_{1/2}\leftrightarrow4D_{5/2}$ transition in $^{86}\text{Sr}^+$ and $^{88}\text{Sr}^+$  with a $1.6\times10^{-11}$ relative  uncertainty to be 570,264,063.435(9) Hz. Furthermore, we detect a relative difference of $3.46(23)\times10^{-8}$ between the orbital g-factors of the electrons in the $4D_{5/2}$ level of the two isotopes. Our method is relatively easy to implement and is indifferent to element or isotope, paving the way for future tabletop searches for new physics and posing interesting prospects for testing quantum many-body calculations and for the study of nuclear structure.
	\end{sciabstract}

	
	In the past several years, precise spectroscopic measurements of atomic and molecular systems have yielded strong constraints on new physics (NP) beyond the Standard Model \cite{Safronova2018}; examples include searches for CPT violations \cite{Kellerbauer2015}, anomalous electron and nucleon electric dipole moments \cite{Engel2013}, local Lorentz invariance breaking \cite{Uzan2011,Pruttivarasin2015}, and time variation of fundamental constants \cite{Engel2013}. Recently, it has been suggested that accurate determinations of atomic ISs can be used to further detect interesting NP candidates \cite{Berengut2018,Delaunay2017}. By testing the linearity of King plot IS comparisons with achievable experimental precision ($\sim 1$ Hz), bounds on the existence of new light force mediators can be improved \cite{Frugiuele2017}.
	
	Naively, reaching such a precision would require measuring optical atomic transitions with a relative uncertainty of $10^{-15}$ or below, a challenging feat accessible only to few leading optical clock labs \cite{Ludlow2015}. Moreover, meaningfully testing the King plot linearity necessitates measuring - at the very least - two optical transitions in four different isotopes, all of which should be known within 1 Hz. Due to these difficulties, state-of-the-art King plot comparisons bound linearity only down to the $\sim$1 kHz range \cite{Drewsen2018}, above the precision necessary for improving existing bounds. 
	
	In quantum metrological scenarios such as the one above, it is often possible to take advantage of techniques that filter out noise but leave the signal intact, thereby increasing sensitivity. A common and powerful technique is that of decoherence free subspaces (DFSs), in which one engineers entangled states that evolve according to target operators, while dynamics take place entirely within a subspace that is invariant (i.e. degenerate) with respect to noise operators. Uses include measurement of the electric quadrupole moment of atoms \cite{roos2006designer}, low uncertainty frequency comparison of two ions \cite{chou2011quantum}, imaging spectroscopy with 100 $\mu Hz$ precision \cite{marti2018imaging}, measurement of the magnetic interaction between electrons in separate atoms at a distance of several micrometers \cite{Kotler2014}, and tests of Lorentz invariance \cite{Pruttivarasin2015}. Employing a DFS for isotope shift measurements was previously suggested by Roos \cite{roos2005precision}, a fact made known to the authors only after completion of the experiment.
	
	In this paper, we present and implement a simple and powerful technique for measuring isotope shifts using a DFS. Our method forgoes measuring the optical transition and instead probes the isotope shift directly. The measurement dynamics take place inside a DFS that is inherently immune to noise terms common to both isotopes, yet sensitive to the IS. Magnetic field and laser-phase noise mitigation allows for a significant prolongation of the measurement coherence time (in our case, a hundredfold), resulting in a corresponding decrease in statistical uncertainty. Immunity to other systematic frequency shifts, such as electric quadrupole, second order Zeeman, and blackbody radiation shifts, entails a low systematic uncertainty budget with comparatively little effort. The method is easy to implement and is not necessarily limited to ions. It essentially requires loading both isotopes into a trap and addressing and measuring each clock transition. We demonstrated our method using $^{88}\text{Sr}^{+}$ and $^{86}\text{Sr}^{+}$ ions and achieved an absolute (relative) uncertainty of 9 mHz ($\Delta f/f \sim10^{-11}$) for this IS measurement. Similar precision with direct measurement of the optical transition frequency would have required a $\sim10^{-17}$ relative uncertainty. We also measured a difference in the orbital angular momentum magnetic susceptibility, $g_L$, between the two isotopes, which we attribute to their small mass difference.

\begin{figure}
		\centering
		\begin{minipage}[b]{.55\textwidth}
			\subfloat[]{\includegraphics[width=\textwidth]{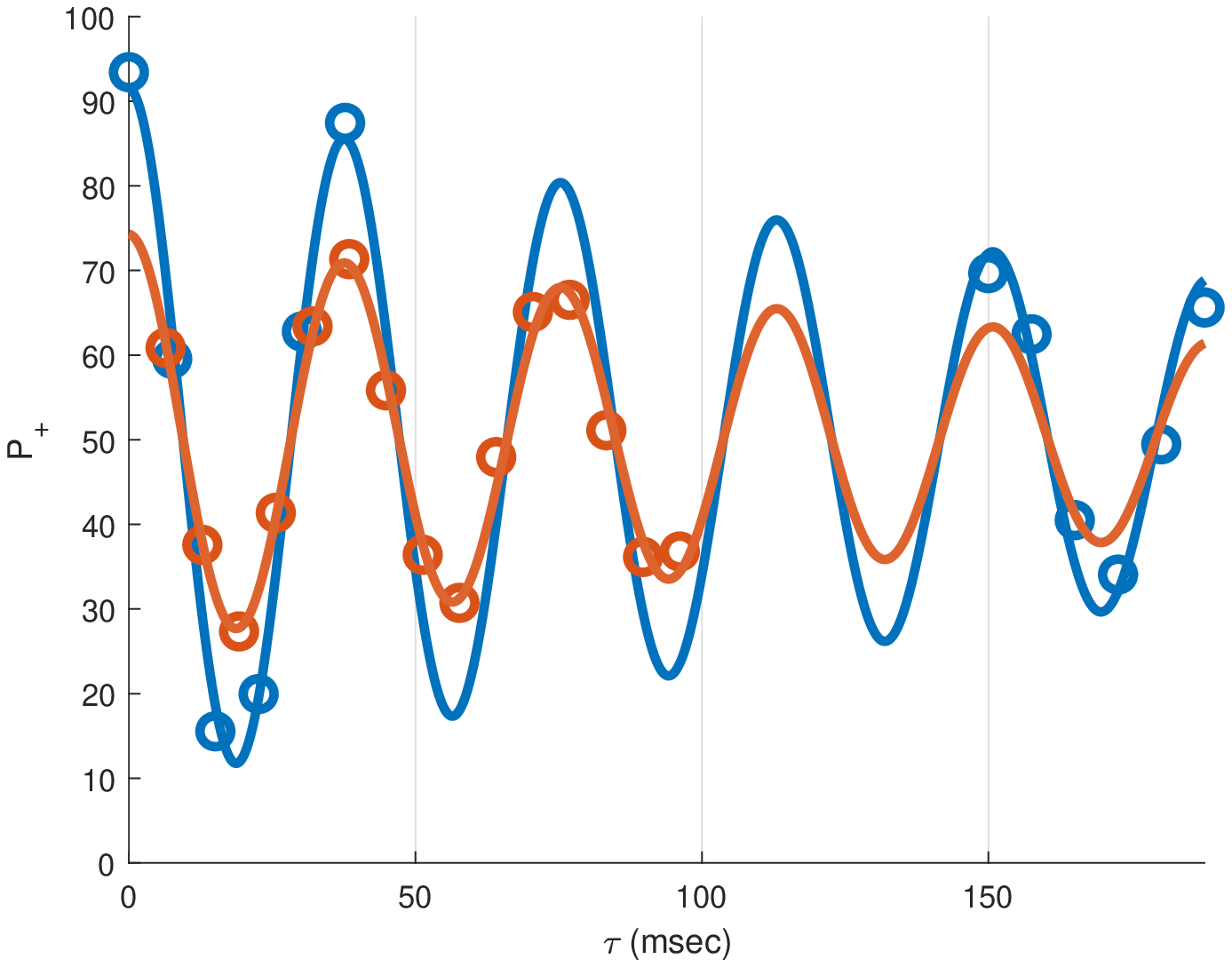} \label{timeScan}}			
		\end{minipage}		
		\begin{minipage}[b]{.35\textwidth}

			\subfloat[]{\includegraphics [height=2.1cm] {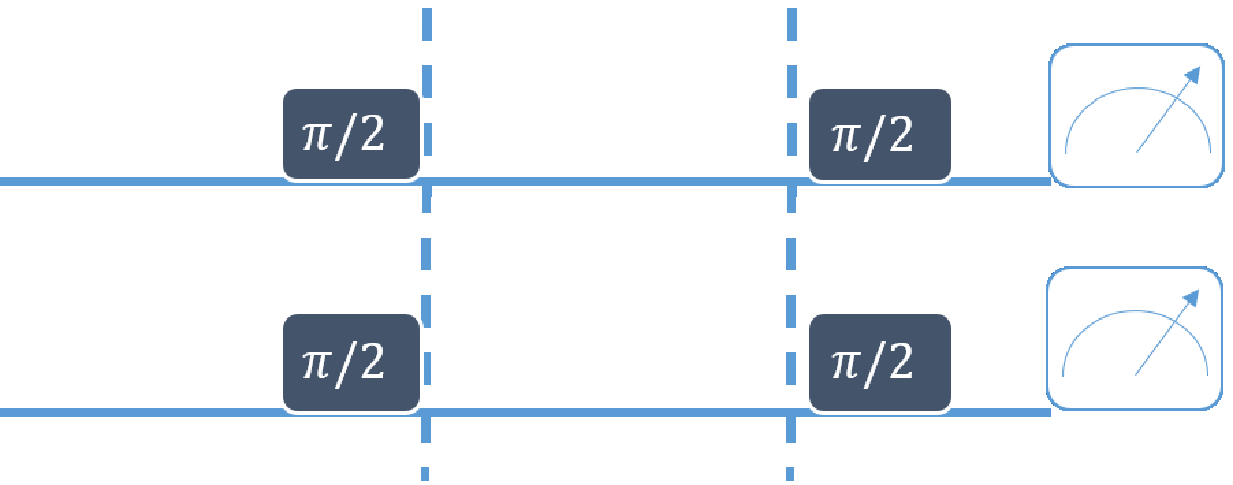} \label{seq_sep}}
			\vfill
			\subfloat[]{\includegraphics [height=2.1cm] {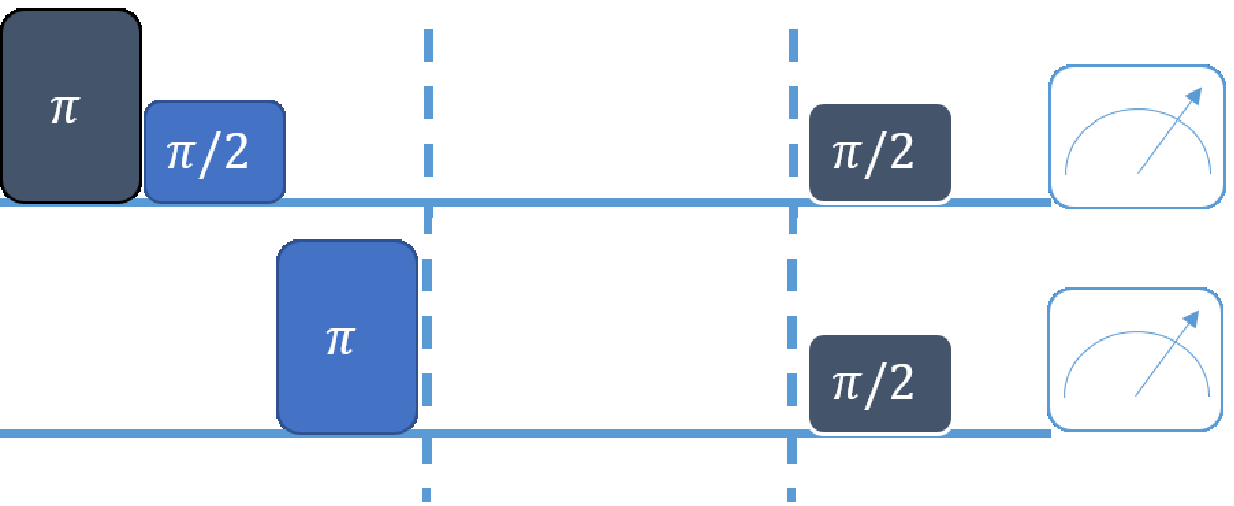} \label{seq_ent}}
			\vfill
		\end{minipage}

		\caption{ (a) The oscillation in time of the parity signal $P_+\equiv P_{ee}+P_{gg}$ for the {\color{MATorange}separable}  experimental sequence shown in (b), and for the {\color{MATblue}entangled} sequence shown in (c); blue-sideband pulses shown in blue, and carrier pulses in black. The contrast of the entangled sequence parity oscillation is almost twice that of the separable sequence, due to the increased correlation of the entangled state. The oscillation frequency is determined by $\delta\nu^{i}_{88,86}-\delta f^{i}_{88,86}$, i.e. the difference between the isotope shift and the addressing fields frequency difference, which in this example is $\sim$26 Hz. Decay is due to the finite lifetime of the excited state. Data is shown in circles, while lines show fit as a guide to the eye. (d) The addressing field difference $\delta f^{i}_{88,86}$ is set by the difference in RF passed into AOMs that modulate a common laser source. The RF difference then serves as a local oscillator, to which the IS is compared.
		}
		
	\end{figure}

\begin{figure}  
		\centering


	\subfloat[]{\includegraphics[width=0.45 \columnwidth]{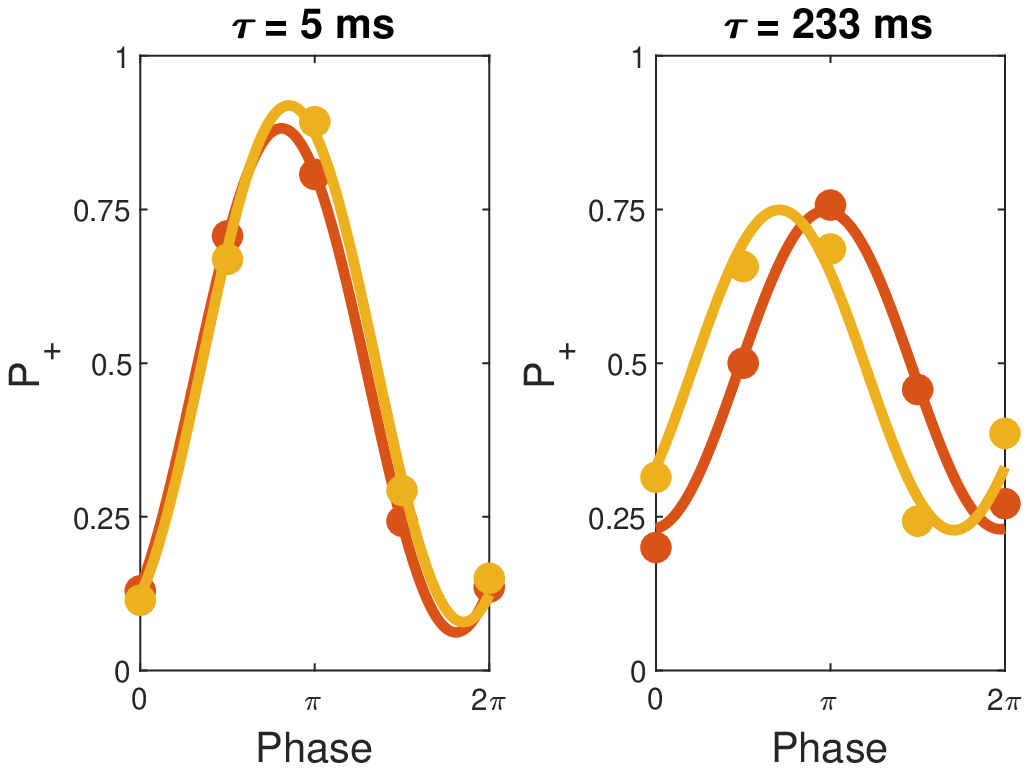} \label{phasescan}}
		\subfloat[]{\includegraphics[width=0.45 \columnwidth]{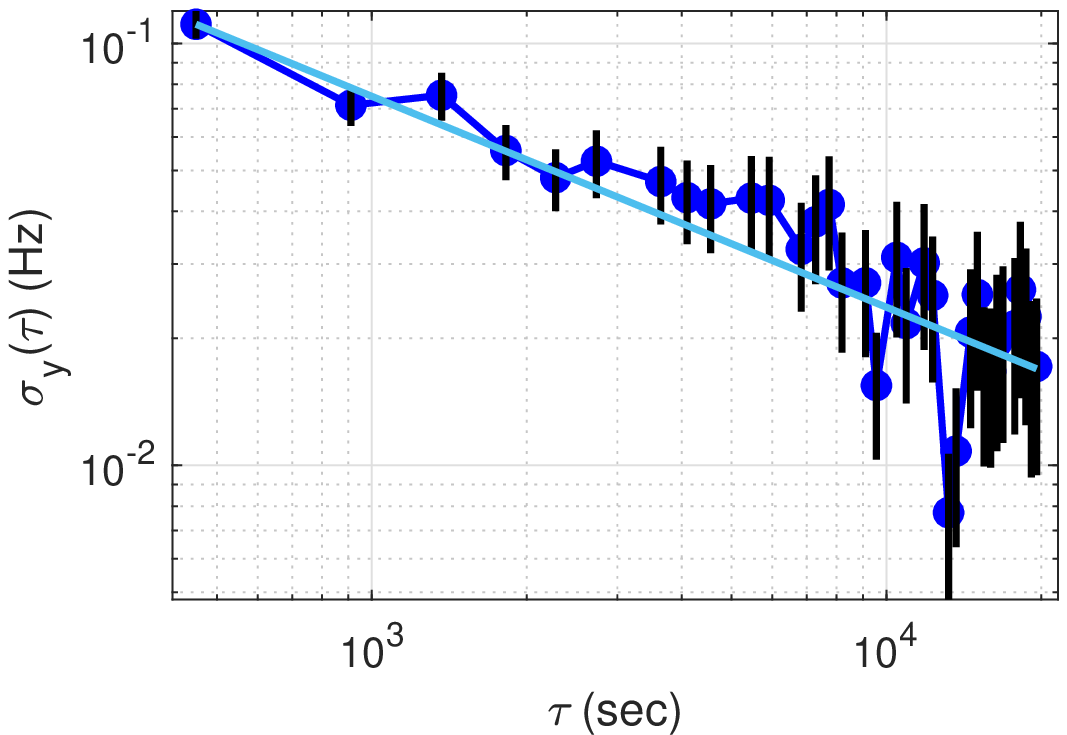} \label{allandev}}\\
		\subfloat[]{\includegraphics[width=0.9 \columnwidth]{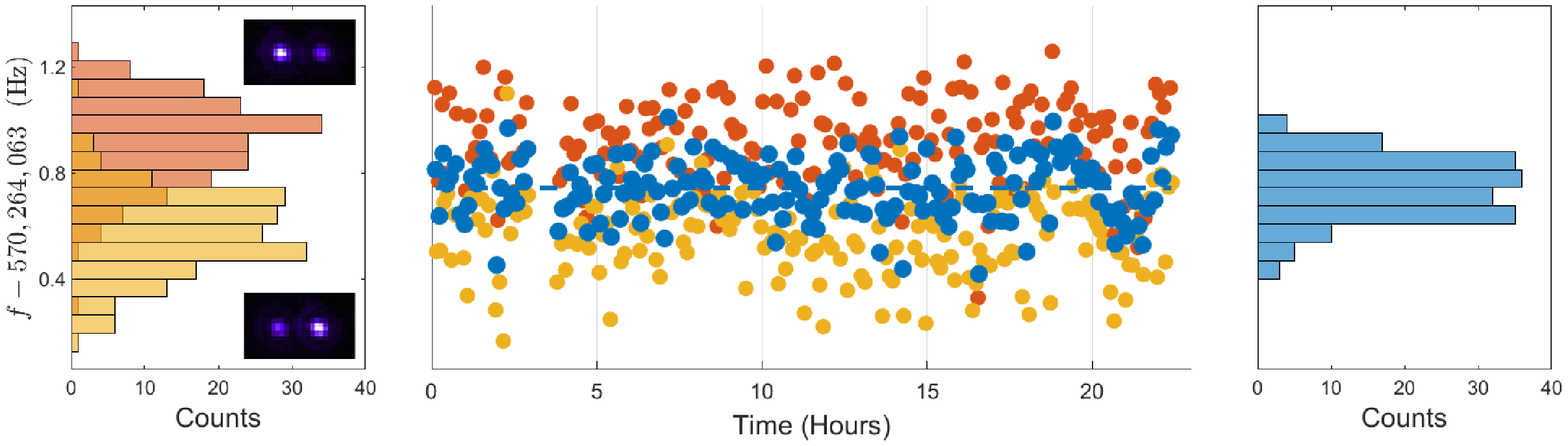} \label{averaging}}
		
		\caption{A 24-hour IS measurement. (a) The superposition phase is determined for short ($\sim5$ ms) and long ($\sim233$ ms) interrogation times by varying the phase difference of the addressing fields and measuring the parity signal $P_{+}=P_{SS}+P_{DD}$. From these the superposition oscillation frequency is extracted. The isotope crystal spatial configuration is alternated in order to average over external field gradients (color coded {\color{MATorange} right} and {\color{MATyellow} left}). (b) Allan deviation analysis of data set shown in (c), consistent with shot-noise limit after 24 hours of averaging. Mean values of every pair of alternate measurements is shown in {\color{MATblue} blue}. This measurement for $m_{S}=1/2$, $m_{D}=3/2$ with an external magnetic field of $5.17$ G yields an oscillation frequency of $570,264,063.745(9)$ Hz.
		}
		\label{experiment}
	\end{figure}

	
	In order to measure the IS; $\delta\nu_{nm}^{i}\equiv\nu_{n}^{i}-\nu_{m}^{i}$; of transition $i$ between states $\ket g$ and $\ket e$, and of isotopes $m,\,n$, we trap a single ion of each isotope in a single chain in a linear Paul trap. Figure \ref{averaging} shows the fluorescence image of such a two-ion crystal. We then prepare the maximally entangled Bell state,
	\begin{equation}
	\ket{\psi_{i}}=\frac{1}{\sqrt{2}}\left(\ket{g_{m}e_{n}}+e^{i\phi_{0}}\ket{e_{m}g_{n}}\right).
	\end{equation} 
	Here $\ket{ge}=\ket g\otimes\ket e$ where the left (right) hand side represents isotope $m$ ($n$) and $\phi_{0}$ is an arbitrary initial phase. This state can be prepared using sideband pulses, as shown in Fig. \ref{seq_ent}. The energy difference between the two states in this superposition, $E_{g_{m}}+E_{e_{n}}-\left(E_{e_{m}}+E_{g_{n}}\right)=\left(E_{e_{n}}-E_{g_{n}}\right)-\left(E_{e_{m}}-E_{g_{m}}\right)=\hbar\left(\nu_{n}^{i}-\nu_{m}^{i}\right)\equiv\hbar\delta\nu_{nm}^{i}$, is exactly the isotope shift times the Planck constant $\hbar$. Therefore, during free evolution for time $\tau$, these states will acquire a relative phase:
	\begin{equation}
	\ket{\psi_{\tau}}=\frac{1}{\sqrt{2}}\left(\ket{g_{m}e_{n}}+e^{i\phi_{0}-i\delta\nu_{nm}^{i}\tau}\ket{e_{m}g_{n}}\right)
	\end{equation} 
	The acquired phase can then be measured and the isotope shift deduced. The Ramsey time $\tau$ is limited only by the coherence time of this superposition, which due to the absence of magnetic field and laser phase noise is determined by the lifetime of the excited states $\ket{e_{m}},\,\ket{e_{n}}$. 
	
	The superposition phase is easily estimated by performing a parity measurement \cite{sackett2000experimental,bollinger1996optimal}. We apply two $\frac{\pi}{2}$ pulses, each at the carrier frequency of one of the isotopes. The phase of each addressing field at time $\tau$ is,
	\begin{equation}
	\varphi_{m,n}= f_{m,n}\tau + \phi_{noise} + \phi_{m,n}.
	\end{equation}
	Here, $f_{m,n}$ are the field frequencies, $\phi_{noise}$ represents laser phase noise which is common to both addressing fields, and $\phi_{m,n}$ denotes an experimentally controllable phase. The resulting populations will obey the relation:
	\begin{equation}
		\begin{gathered}
				p_{ee}+p_{gg}-\left(p_{eg}+p_{ge}\right)=\cos\left(\phi_{0}+\delta\nu_{nm}^{i}\tau-\left(\varphi_{m}-\varphi_{n}\right)\right)\\
				= \cos\left(\phi_{0}+(\delta\nu_{nm}^{i}-\delta f_{nm})\tau-\left(\phi_{m}-\phi_{n} \right)\right)
		\end{gathered}
	\end{equation}\\
	The parity signal will oscillate in time at the detuning of the two fields' frequency difference, which can be dictated by an RF source, with respect to the isotope shift. Laser phase-noise is cancelled since it is common to both addressing fields.
	
	A simpler version of this method, yet with lower signal-to-noise ratio,  can be implemented without the need for entanglement. Here, we prepare each isotope individually in an equal superposition and arrive at the two-ion state,
	\begin{equation}
	\ket{\psi_{0}^{\prime}}=\frac{1}{2}\left(\ket{g_{m}g_{n}}+\ket{g_{m}e_{n}}+\ket{e_{m}g_{n}}+\ket{e_{m}e_{n}}\right).
	\end{equation}  
	After a time $\tau_{1}\gg\tau_{d}$, where $\tau_{d}$ is the dephasing time due to laser phase noise and magnetic field fluctuations, the states $\ket{g_{m}g_{n}}$ and $\ket{e_{m}e_{n}}$ will completely dephase, resulting in the density matrix $\rho_{\tau_{1}}=\frac{1}{4}\left(\ketbra{g_{m}g_{n}}+\ketbra{e_{m}e_{n}}\right)+\frac{1}{2}\ketbra{\psi_{\tau_{1}}}$. Hence, with probability $\frac{1}{2}$ the required state is generated, and with probability $\frac{1}{2}$ the phase information is lost. The dephased part will average to a phase-insensitive null background in the parity measurement, which will now yield:
	\begin{equation}
		\begin{gathered}
			p_{ee}+p_{gg}-\left(p_{eg}+p_{ge}\right)=\\
			\frac{1}{2}\cos\left(\phi_{0}+(\delta\nu_{nm}^{i}-\delta f_{nm})\tau-\left(\varphi_{m}-\varphi_{n}\right)\right).
		\end{gathered}		
	\end{equation}
	The cost of dephasing is a reduced contrast of the Ramsey fringe, corresponding to a loss of half the signal. This emphasizes the fact that here we are measuring correlations between two atoms, where the entanglement simply serves to prepare a maximally correlated state. 
	
Measurements of the IS of the electron orbitals are, in general, affected by the weak dependence of the magnetic susceptibility of the orbital angular momentum on the mass of the nucleus. For a Hydrogen-like atom, the orbital g-factor is corrected to be $g_{L}\approx1-m_{e}/m_{N}$ where $m_{e}$ is the electron mass and $m_{N}$ is the nucleus mass \cite{Lamb1952}. This effect is analogous to the normal mass shift of electron orbitals. For a many-electron atom, an additional correction due to correlations between electrons appears (see Supplementary), analogous to the electron orbital specific mass shift  \cite{Bartlett1933}. The differential magnetic susceptibility can be measured and eliminated by comparing measurements of different or opposite excited $m$ levels and at different magnetic fields.

	In our experiment, we measure the IS of the narrow (0.4 Hz)  $\ket{5S_{\frac{1}{2}}}\leftrightarrow\ket{4D_{\frac{5}{2}}}$ electric-quadrupole transition between $^{88}\text{Sr}^{+}$ and $^{86}\text{Sr}^{+}$. Both clock levels are split by a DC magnetic field of 3-5G to their Zeeman sub-levels which are separated by several MHz. We trap both isotopes in a linear Paul trap with 1.3 MHz axial center-of-mass and 3.5 MHz radial harmonic frequencies. State-selective fluorescence detection, state initialization, and Doppler cooling are performed by illuminating the ions with bichromatic laser fields, resonant with the relevant dipole-allowed transitions of both isotopes, enabling non-ambivalent preparation and readout of the two ion state. 
	
	The $5S_{1/2}\rightarrow4D_{5/2}$ transition is driven by a narrow ($\sim$20 Hz) linewidth laser \cite{peleg2019phase}. In order to resonantly address both isotopes, we split the laser into different AO frequency shifters that bridge the $\sim$570 MHz IS gap. The two beams are recombined and sent through another AO frequency shifter for common frequency control, and then passed to the ions through a single mode fiber to minimize differential optical phases along different paths. 
		
	The ions are initially Doppler cooled and optically pumped to a chosen Zeeman sub-level of the $5S_{1/2}$ manifold. They are then sideband-cooled to the ground state of the axial center-of-mass mode, via the $^{88}\text{Sr}^{+}$ ion, yielding the state $\ket{S_{88}S_{86}0}$, where letters denote electron orbitals, subscripts denote isotopes and the last number denotes the excitation level of the axial center of mass motional mode. The entangled state is then prepared using the following sequence. First, a carrier $\pi$ pulse on the $^{88}\text{Sr}^{+}$ generates the state $\ket{D_{88}S_{86}0}$. Then, a $\pi/2$ pulse on the blue sideband transition of the $^{86}\text{Sr}^{+}$ ion generates the state \\$\frac{1}{\sqrt{2}}\left(\ket{D_{88}S_{86}0}+e^{i\phi_{0}^{\prime}}\ket{D_{88}D_{86}1}\right)$, where $\phi_{0}^{\prime}$ is some arbitrary phase. Finally, a blue sideband $\pi$ pulse on the $^{86}\text{Sr}^{+}$ ion produces the state $\frac{1}{\sqrt{2}}\left(\ket{D_{88}S_{86}}+e^{i\phi_{0}}\ket{D_{86}S_{88}}\right)\otimes\ket 0$ with $\phi_{0}$ as some arbitrary initialization phase. Here, $D$ and $S$ represent particular Zeeman sub-levels of the corresponding manifolds. We perform entangled state initialization with a fidelity of ~0.8-0.9, limited by the quality of the blue sideband pulses, which is in turn limited by the quality of the sideband cooling and the ion heating rate.

	Following a free evolution time $\tau$, the phase difference between the closing $\frac{\pi}{2}$ pulses is scanned from 0 to 2$\pi$, giving a parity signal with some phase $\phi$. This is repeated for two different times $\tau_{i}<\tau_{f}$. We extract the IS frequency from the phase and time differences between the two sequences. As gradients of external fields, most prominently a gradient of the magnetic field, can generate an additional frequency difference between the two ions, we repeat the measurement for a reversed order of the two different isotopes in the crystal (fig. \ref{phasescan}). The average frequency of these two measurements is then added to our local oscillator frequency which is given by the difference of RF frequencies controlling the lasers for the isotopes. All RF sources are locked to a GPS-referenced high quality oscillator (MenloSystems GPS 8-12) stabilized within $5\times10^{-12}$ in one second. Averaging over many repetitions of this measurement gives an isotope shift for the transition under some choice of the Zeeman levels $m_{S}$, $m_{D}$ and the magnetic field B, which is measured independently. In order to determine the isotope shift at zero magnetic field, this process is repeated for Zeeman levels -$m_{S}$, -$m_{D}$, and at a different B. Combining all measurements we determine $\delta g_{L}$ and $\delta\nu_{88,86}^{SD}$ by a maximum-likelihood analysis. 
	
	
	We perform three sets of measurements, defined by specific combinations of external magnetic field B and Zeeman levels: $B\approx5.17$ G, $m_S=1/2$, $m_{D}=3/2$; $B\approx3$ G, $m_S=1/2$, $m_{D}=3/2$; and $B\approx3$ G, $m_S=-1/2$ $m_{D}=-3/2$. The external field is measured independently by spectroscopy of the transition between the two Zeeman states of the S manifold. For every set, we measure 2-4 traces, each consisting of 12-24 hours of averaging (example shown in fig. \ref{averaging}). Traces of the same set are measured as far as three weeks apart demonstrating the system's stability. We calculate the Allan deviation (fig. \ref{allandev}) for each trace and find all traces to be shot noise limited; statistical uncertainty for a single trace varies from 9 to 15 mHz standard deviation. All traces are consistent with an electronic IS as well as an IS of the excited state magnetic susceptibility $g_{D}$, showing a linear dependence on the external magnetic field.

	\begin{figure}
		\centering
		\subfloat[]{\label{ISvB}\includegraphics[width=0.48 \columnwidth]{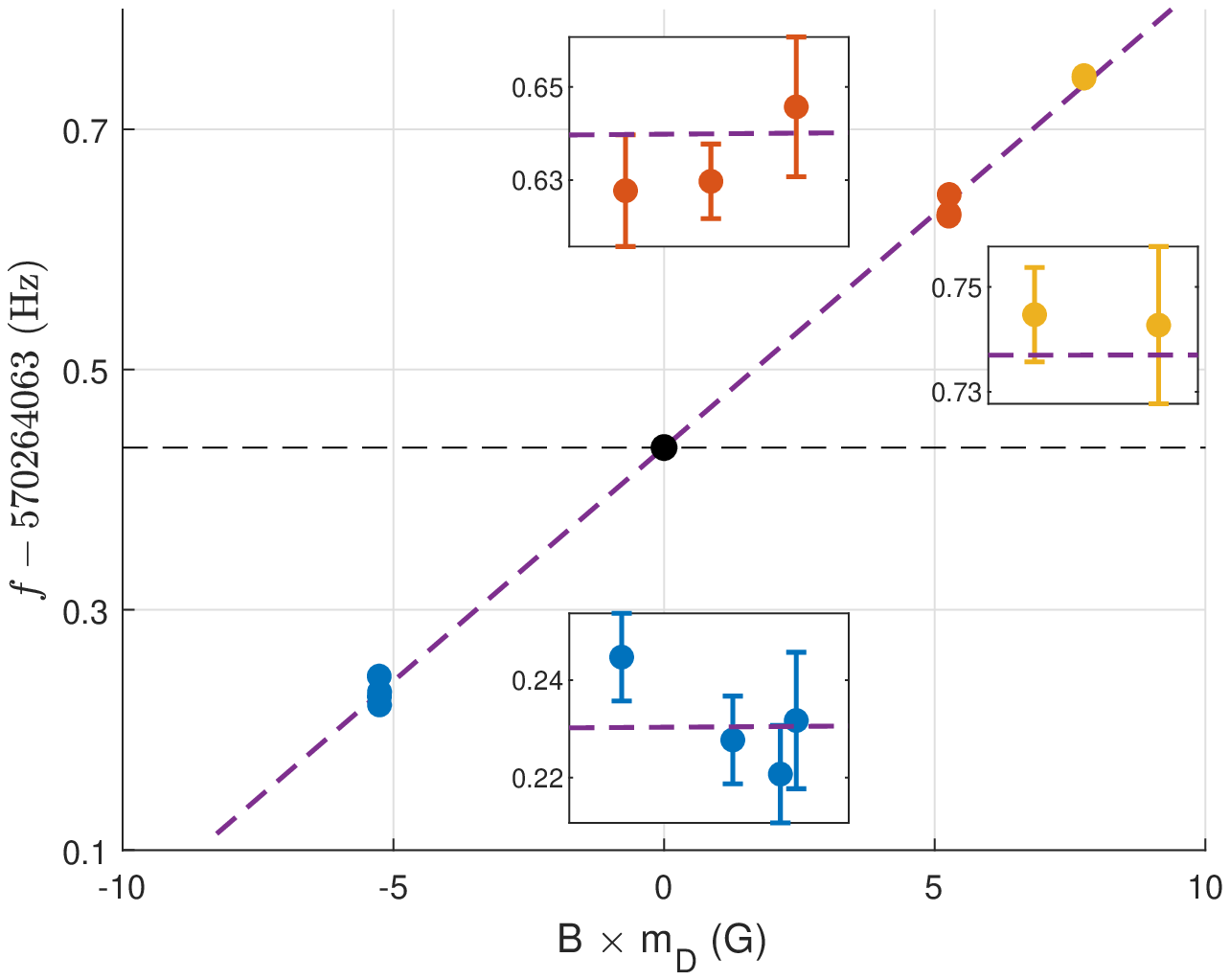} }
		\subfloat[]{\label{ISvT}\includegraphics[width=0.48 \columnwidth]{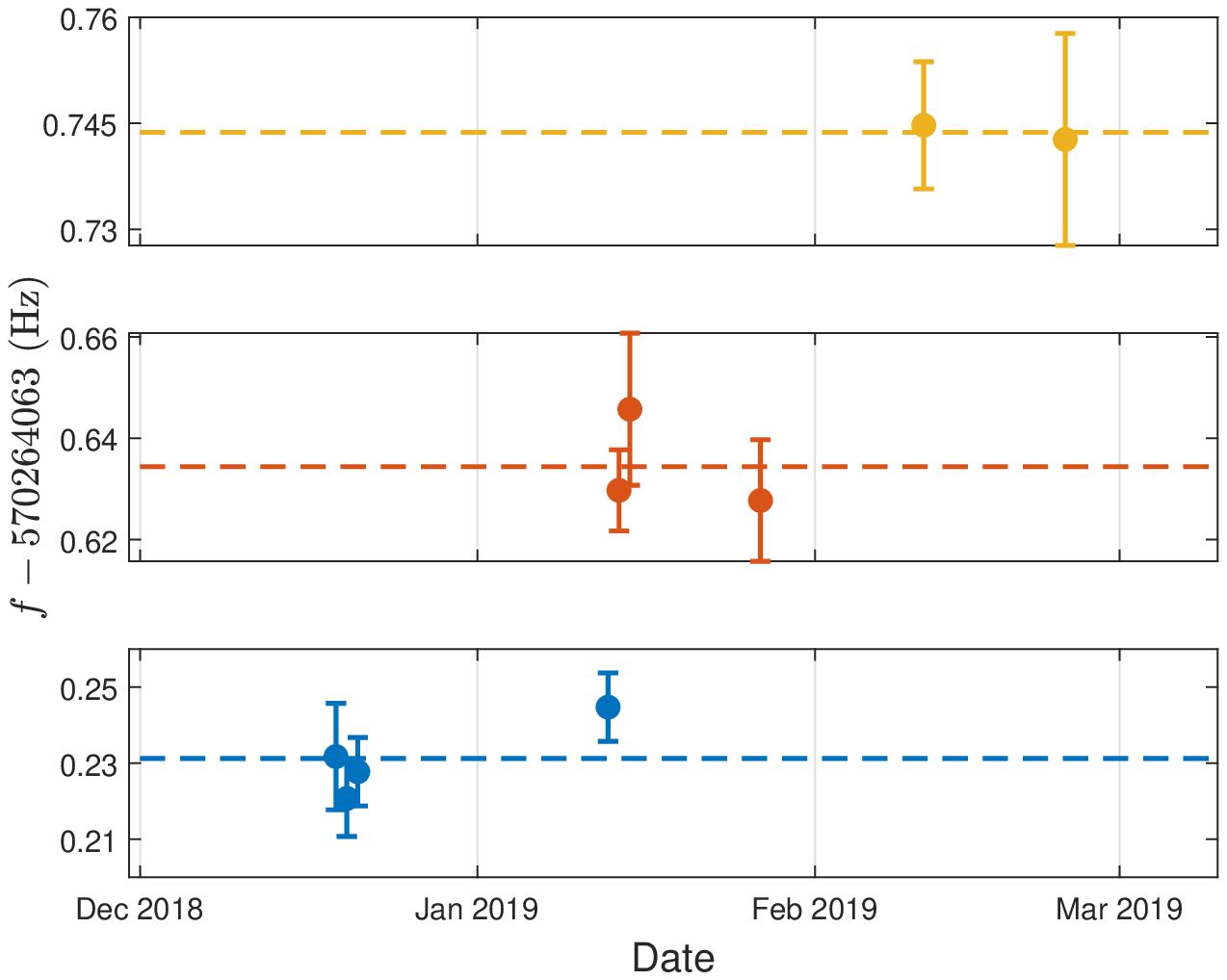} }

		\caption{Results summary: measuring $\delta\nu^{SD_{5/2}}_{88,86}$ and $\delta g_{D}^{88,86}$. (a) IS as a function of $B\times m_{D}$. Colors correspond to {\color{MATblue}$B\approx3.5$G, $m_D=-3/2$}; {\color{MATorange}$B\approx3.5$G, $m_D=3/2$}; and {\color{MATyellow}$B\approx5.17$G, $m_D=3/2$}. Maximum likelihood linear fit shown in {\color{MATpurple} purple}. Standard deviation confidence intervals shown in insets. The IS variation in $B\times m_D$ is due to $g_D$ dependence on isotope, of which the slope of 0.0388(26) Hz/G is a direct measurement. The black dot denotes the IS at null magnetic field: 570,264,063.435(9) Hz. (b) IS over time. Measurements taken several weeks apart remain within the statistical uncertainty of a single measurement. Error bars represent statistical uncertainty only for all plots. 
		}
		
	\end{figure}

	The main systematic shifts for optical ion-clocks measuring at $<1$ Hz uncertainty include the electric quadrupole shift, the blackbody radiation (BBR) shift, the 2nd order Zeeman shift, and shifts due to the trap RF: an AC Stark shift and a 2nd-order Doppler shift \cite{Ludlow2015}. BBR, trap AC Stark shifts and quadrupole shifts are common mode rejected up to the small isotopic changes to the electronic wavefunction ($\sim10^{-6}$), rendering them negligible ($<10^{-4}$ Hz). The differential second-order Zeeman shift is similarly negligible. Doppler shifts are only cancelled up to the mass difference between the ions, which provides a $1/40$ reduction; however, as these shifts are relatively small to begin with ($\sim10$ mHz), the mass reduction is sufficient to render them negligible as well.
	
	Uniquely to our scheme, gradients of external fields can induce slow noise or constant systematic shifts. We mitigate their effect considerably by alternating the isotope positions and averaging over the measured shifts. However, the isotope position exchange is imperfect due to the mass dependence of the RF pseudo-potential. Specifically, as the RF gradient is pronounced along the radial direction, isotopes that have exchanged axial positions can retain some radial offset due to uncompensated radial DC fields and the mass difference. In this case, fields that have a non-negligible radial gradient over the offset distance, e.g. the external magnetic field and the RF field itself, will result in a systematic shift. The magnitude and direction of the offset, and therefore also the systematic shift, will depend on the quality of RF micromotion compensation, as when micromotion is compensated the ions are trapped exactly along the RF null and uncompensated radial dc fields are minimized.
	
	We restrict the effect of both the residual magnetic field and RF gradients on the IS by independently measuring each. We determined the RF amplitude by addressing micromotion sidebands in three dimensions, from which we calculated the combined effect of AC Stark and 2nd-order Doppler shifts \cite{Dube2013}. The magnetic field effect was measured directly by repeating a variation of the IS experiment using the spin states of the $5S_{1/2}$ Zeeman manifold; as $\Delta g_{S}$ is very small \cite{Koehler2016}, a non-zero IS measurement in this manifold indicates the existence of a radial magnetic field gradient term. We found that the residual RF field difference contributes a systematic shift which we bound at 2 mHz, and the magnetic field radial gradient similarly contributes a systematic shift which we bound at 8 mHz, on par with our statistical uncertainty of 5 mHz. In future experiments this systematic shift can be mitigated by minimizing magnetic gradients in all directions. In addition, using these measurements we place an upper bound on the isotope shift of the electron spin g-factor, at $\frac{\lvert\delta g_S^{88,86}\rvert}{g_S}<2.27\times10^{-9}$ with 95\% confidence.
	
	The final results of our measurement are shown in \ref{ISvB}. Overall, we measure ten averaging traces in three measurement sets. Plotting the measured IS as a function of $B\times m_L$, we obtain a linear relation corresponding to a pure electronic isotope shift and a magnetic susceptibility isotope shift, generating the slope. From a maximum likelihood fit to a linear relation, we obtain an isotope shift at null magnetic field of $\delta\nu^{S,D}_{88,86}$= 570,264,063.435(9)(5)(8) Hz (total)(statistical)(systematic), which corresponds to a relative uncertainty of $1.6\times10^{-11}$. Our uncertainty is a $\sim10^{-17}$ fraction of the optical transition frequency. We also measure a difference in susceptibility of 0.0388(26) Hz/G, corresponding to a relative susceptibility isotope shift of: $\frac{\delta g_D^{5/2}}{g_D^{5/2}}=\frac{0.0388(26)}{1.68\times10^6}=2.31(15)\times10^{-8}$, which translates to a relative susceptibility shift of the orbital angular momentum $\frac{\delta g_L}{g_L}=3.46(23)\times10^{-8}$.

	
	Our experiment highlights several noise sources that must be considered in future attempts to measure ISs at $<1$ Hz resolution. The most prominent of these is the IS of the magnetic susceptibility, which can easily induce shifts on the order of several Hz. For instance, the normal mass susceptibility shift for the $^{48,40}\text{Ca}^{+}$ isotopes is $\sim2.2\times10^{-6}$, which would translate to a $\sim2.5$ Hz shift per Gauss per angular momentum quanta (with our experimental parameters, an overall shift of $\sim20$ Hz).  As we show, this effect can be simply dealt with by measuring opposite Zeeman states. Furthermore, our experiment demonstrates that in order to reach the $<1$ mHz regime, careful characterization and monitoring of micromotion is essential, as well as high quality suppression of magnetic field gradients in all directions.

	The ability to interrogate optical isotope shifts with substantially improved precision presents a host of new opportunities for nuclear physics research. Due to their sensitivity to the nuclear charge radius, optical IS measurements are routinely used in order to probe nuclear structure \cite{Ruiz2016,Mueller2007,Sanchez2006}; in fact, optical IS measurements provide the most stringent bounds on nuclear charge radii for a large number of isotopes \cite{Kluge2003,Angeli2013}. Moreover, for heavy nuclei, optical IS measurements are also sensitive to higher order moments of the nuclear charge distribution \cite{Angeli2013}.

	Precision IS measurements can also serve as a challenge and testbed for many-body atomic calculations. The specific mass shift (SMS) term requires evaluating the many-body electron correlation $\langle\sum_{ij}\vec{p_i}\cdot\vec{p_j}\rangle$, which is known to be difficult to calculate \cite{Berengut2003,Safronova2001}. Using our method, the SMS can be resolved with an accuracy beyond what was available so far \cite{Gebert2015}. The $^{48,40}\text{Ca}^{+}$ IS, for instance, will be highly dominated by the mass shift (as both isotopes are doubly magic, canceling field shift effects), allowing a direct high accuracy measurement. Beyond the SMS, we are able to measure the many-body correlation term $\langle\sum_{ij}\vec{r_i}\times\vec{p_j} \rangle$ through the specific orbital magnetic susceptibility, providing yet another independent test of the many-body wave function (for more details see Supplementary Information). This term is evaluated directly in the g-factor IS measurement, as here there is no contribution to the nuclear finite charge radius complicating the analysis.
	Our isotope shift scheme can also be used to investigate the isotope dependence of $g_S$, the magnetic susceptibility of the bound electron, by using the state $\frac{1}{\sqrt{2}}(\ket{\uparrow\downarrow}+\ket{\downarrow\uparrow})$. This state has no inherent limit to coherence, and one can maintain the superposition for tens of seconds or more \cite{Kotler2014}. In ion traps, the coherence time for such a state is often limited by ion heating rates \cite{Kotler2014}, but even this limitation can be overcome by using cryogenic traps or sympathetic cooling. Consequently, if systematic noise sources such as magnetic field gradients are properly treated, the long coherence time can potentially allow one to detect extremely small effects, far under 1 mHz. By using slightly stronger magnetic fields, $g_S$ can possibly be evaluated at $\frac{\delta g_S}{g_S}<10^{-11}$, paving the way for tests of QED and possibly beyond SM effects \cite{Koehler2016}. 
	
	The two-isotope entangled superposition can be thought of as a synthetic RF IS atomic clock. Clocks use a stable periodic phenomenon in nature in order to keep track of time by counting periods. An IS clock is conceptually unique in the sense that, instead of using a local periodic phenomena, it uses non-local correlations between two oscillating subsystems as a periodic reference. Despite the instability of each subsystem the correlations remain highly stable and can serve as a reliable reference with low systematic uncertainty.
	
	In summary, we present and demonstrate a novel method for measuring isotope shifts with trapped ions and use this method to measure the isotope shift between $^{88,86}\text{Sr}^{+}$ with high precision. The method makes use of the existence of a decoherence free subspace which is invariant to the most dominant noises, both fast and systematic. The method is simple and easy to use, as in essence it requires no more than carrier pulses and measurement of two isotopes. Beyond measuring the IS of the optical transition, we determine the IS of the orbital magnetic susceptibility, which is sensitive to many-body electron correlations. Precision IS measurements open new possibilities in nuclear and atomic physics. Besides Paul traps, our method can be applied to metrology with optical tweezers \cite{norcia2019seconds}. The precision we demonstrate is far better than needed in order to potentially bound beyond SM physics (by repeating the experiment with several isotopes), and is the most precise optical IS measurement to date in terms of relative uncertainty \cite{takano2017precise}. 

\paragraph{Acknowledgements---}
This work was supported by the Israeli Science Foundation, the Israeli Ministry of Science Technology and Space, the Minerva Stiftung and the European Research Council (consolidator grant 616919-Ionology).
\pagebreak

\bibliographystyle{ieeetr}
\bibliography{bib_experiment_a}



\pagebreak	
\section{Supplementary}

	\subsection{g-factor IS}	 	
 	A Hydrogen-like atom should have a leading-order nuclear recoil correction to $g_{L}$ of $-m/M$. For the $D_{5/2}$ level in Sr, this would manifest as a correction to $g_{J}=0.8g_{L}+0.2g_{S}$, where $\frac{\delta g_{J}}{g_{J}}\approxeq\frac{0.8}{1.2}(\frac{m_{e}}{m_{86}}-\frac{m_{e}}{m_{88}})\approxeq9.64\times10^{-8}$. This calculated result is larger than (although on the same order of magnitude as) the measured correction of $2.31(15)\times10^{-8}$. We suggest that in a manner similar to that of the optical IS, the discrepancy is due to the added energy shift of electron correlations termed the "specific" mass shift, which in general is on the same order of magnitude as the normal mass shift. In fact, the IS of the magnetic susceptibility may serve as a direct probe of these correlations and thus of the electron many-body wave function.
 	
 	We derive the specific mass orbital susceptibility term. In the center of mass frame, $\vec{p_n}=-\sum_{i}\vec{p_i}$ and $\vec{r_n}=-\frac{m}{M}\sum_{i}\vec{r_i}$, where $\vec{r_n},\vec{p_n}$ and $\vec{r_i},\vec{p_i}$ are the nuclear and electronic coordinates, respectively. The nuclear angular momentum is:
 	\begin{align*}
	 	\vec{L_n}=\vec{r_n}\times\vec{p_n}&=\frac{m}{M}\sum_i\vec{r_i}\times\sum_i\vec{p_i}\\
	 	&=\frac{m}{M}(\sum_i\vec{r_i}\times\vec{p_i}+\sum_{j\neq k}\vec{r_j}\times\vec{p_k})\\
	 	&=\frac{m}{M}(\sum_i\vec{L_i}+\sum_{j\neq k}\vec{r_j}\times\vec{p_k})
 	\end{align*}
	So the total angular momentum can be written as:
	\begin{equation} 
		\label{totalL}
		\vec{L}=\sum_{i}\vec{L_{i}}+\vec{L_n}=(1+\frac{m}{M})\sum_i\vec{L_i}+\frac{m}{M}\sum_{j\neq k}\vec{r_j}\times\vec{p_k}
	\end{equation} 
	The Zeeman Hamiltonian is:
	\begin{equation}
		\label{ZeemanHam}
		H_{B}=\vec{B}\cdot(-\frac{e}{2m}\sum_{i}\vec{L_{i}}+\frac{e}{2M}\vec{L_n})\approx	\vec{B}\cdot(-\frac{e}{2m}\sum_{i}\vec{L_{i}})
	\end{equation}
	Isolating the electronic term in \eqref{totalL}, 
	\begin{align*}
	\sum_i\vec{L_i}&=\frac{\vec{L}-m/M \sum_{j\neq k}\vec{r_j}\times\vec{p_k}}{1+m/M}\\
	&\approxeq\vec{L}-\frac{m}{M}\vec{L}-\frac{m}{M}\sum_{j\neq k}\vec{r_j}\times\vec{p_k}
	\end{align*}	
	we can rewrite the Hamiltonian \eqref{ZeemanHam} as:
	\begin{equation}
	H_B=\frac{-e}{2m}\vec{B}\cdot(\vec{L}-\frac{m}{M}\vec{L}-\frac{m}{M}\sum_{j\neq k}\vec{r_j}\times\vec{p_k})
	\end{equation}
	and choosing $\vec{B}=B\hat{z}$ we get:	
	\begin{equation}
		\mu_z=\frac{-e}{2m}[L_z(1-\frac{m}{M})-\frac{m}{M}\sum_{j\neq k}(\vec{r_j}\times\vec{p_k})_z]
	\end{equation}
	The second term in parentheses is the Hydrogen-like contribution, corresponding to the normal MS. The third term is exclusive to many-electron atoms and depends on interelectronic correlations, and is thus a specific MS. In order to show that the third term is indeed a susceptibility, one must show that its expectation value is (at least to first order) linear in $m_z$, i.e. that
	\begin{equation}
	\bra{\alpha_1\dots\alpha_n,j,m}(\vec{r_j}\times\vec{p_k})_z\ket{\alpha_1\dots\alpha_n,j,m}\approx m_z\cdot f(\alpha_1\dots\alpha_n) 
	\end{equation}
	
	This can shown to hold directly from the Wigner-Eckart theorem. The many body operator $\vec{L_n}=(\sum_i\vec{r_i})\times(\sum_j\vec{p_j})$ is a vector operator, and hence its components can be written in terms of spherical tensor operators $\hat{T}^{1}_{q=0,\pm 1}$, where the $\hat{z}$ component is simply $\hat{T}^1_0$. Hence
	\begin{align*}
	&\bra{\alpha_1,\dots\alpha_n,j,m}(\sum_i\vec{r_i}\times\sum_j\vec{p_j})_z\ket{\alpha_1,\dots\alpha_n,j,m}=\\ &\braket{j,m,1,0}{j,m}\langle\alpha_1,\dots\alpha_n,j\parallel\hat{T}^1\parallel\alpha_1,\dots\alpha_n,j\rangle=\\
	&\frac{m}{\sqrt{j(j+1)}}\langle\alpha_1,\dots\alpha_n,j\parallel\hat{T}^1\parallel\alpha_1,\dots\alpha_n,j\rangle \numberthis \label{eqn}
	\end{align*}
	
	Where the bracketed term is a radial part which is independent of $m$, but generally dependent on all other quantum numbers. Hence, the mass shift of the orbital angular momentum susceptibility can be divided into an normal term $k_{NMS}$ and a specific term $k_{SMS}$:
	\begin{align*}
	&g_{L} = 1-k_{NMS}-k_{SMS}\\
	&k_{NMS}=\frac{m}{M}; \\ &k_{SMS}=\frac{m}{M}(\frac{\langle\alpha_1,\dots\alpha_n,j\parallel\hat{T}^1\parallel\alpha_1,\dots\alpha_n,j\rangle}{\sqrt{j(j+1)}}-1)
	\end{align*}

\end{document}